\documentclass[aps,prl,reprint,superscriptaddress,twocolumn]{revtex4-1}        

\def\fm#1{\ifmmode #1 \else $#1$\fi}

\usepackage{xspace}
\def\ra{\fm{\rightarrow}\xspace}

\newcommand{\captionmacro}[2][]{\caption[#1]{\textbf{#1} #2}}

\usepackage{multirow}
\usepackage{graphicx}
\usepackage[normalem]{ulem}

\usepackage{amsmath}

\usepackage{booktabs}  
\AtBeginDocument{
\heavyrulewidth=.08em
\lightrulewidth=.05em
\cmidrulewidth=.03em
\belowrulesep=.65ex
\belowbottomsep=0pt
\aboverulesep=.4ex
\abovetopsep=0pt
\cmidrulesep=\doublerulesep
\cmidrulekern=.5em
\defaultaddspace=.5em
}

\RequirePackage{siunitx}
\def\szone{\fm{S}\xspace}

\def\ezone{\fm{A}\xspace}
\def\epzone{\fm{B}\xspace}
\def\czone{\fm{C}\xspace}
\def\cyzone{\fm{C'}\xspace}

\def\hzone{\fm{H}\xspace}
\def\vzone{\fm{V}\xspace}

\def\Sab{\fm{S_{ab}}\xspace}
\def\AaBb{\fm{A_aB_b}\xspace}
\def\SaRb{\fm{S_aR_b}\xspace}
\def\LaSb{\fm{L_aS_b}\xspace}
\def\BaAb{\fm{B_aA_b}\xspace}
\def\Sba{\fm{S_{ba}}\xspace}

\def\SuaRb{\fm{\textit{\underline{S}}_aR_b}\xspace}
\def\LaSub{\fm{L_a\textit{\underline{S}}_b}\xspace}

\def\AiBj{\fm{A_iB_j}\xspace}
\def\SiRj{\fm{S_iR_j}\xspace}
\def\LiSj{\fm{L_iS_j}\xspace}

\def\Sa{\fm{S_{a}}\xspace}
\def\Sua{\fm{\textit{\underline{S}}_{a}}\xspace}

\def\Aa{\fm{A_{a}}\xspace}
\def\Ca{\fm{C_{a}}\xspace}
\def\Cya{\fm{C'_{a}}\xspace}
\def\Ha{\fm{H_{a}}\xspace}
\def\Va{\fm{V_{a}}\xspace}

\def\Suab{\fm{\textit{\underline{S}}_{ab}}\xspace}

\usepackage{siunitx}

\usepackage{braket}

\def\be{${^\mathrm{9}}$Be$^\mathrm{+}$\xspace}

\usepackage{todonotes}

\usepackage{cleveref} 
\crefname{figure}{Fig.}{Figs.}
\crefname{table}{Table}{Table}
\crefname{equation}{Eq.}{Eqs.}
\crefname{chapter}{Chapter}{Chapters}
\crefname{section}{Sec.}{Sec.}
\crefname{appendix}{Appendix}{Appendices}
\crefname{appchap}{Appendix}{Appendices}
\crefname{appsec}{Appendix}{Appendices}

\begin{document}        


\title{Ion transport and reordering in a two-dimensional trap array}


\author{Y. Wan}
\email{wanyong@nim.ac.cn}  
\altaffiliation[Current address: ]{National Institute of Metrology, 18 Changchi Road, Changping District, Beijing 102200, China}
\affiliation{National Institute of Standards and Technology,  Boulder, CO 80305, USA}
\affiliation{Department of Physics, University of Colorado,	Boulder, CO 80309, USA}

\author{R. J\"ordens}
\altaffiliation[Current address: ]{QUARTIQ GmbH, Berlin, Germany}
\affiliation{National Institute of Standards and Technology, Boulder, CO 80305, USA}
\affiliation{Department of Physics, University of Colorado,	Boulder, CO 80309, USA}

\author{S.~D.~Erickson}
\affiliation{National Institute of Standards and Technology, Boulder, CO 80305, USA}
\affiliation{Department of Physics, University of Colorado,	Boulder, CO 80309, USA}

\author{J.~J.~Wu}
\affiliation{National Institute of Standards and Technology, Boulder, CO 80305, USA}
\affiliation{Department of Physics, University of Colorado,	Boulder, CO 80309, USA}


\author{R. Bowler}
\altaffiliation[Current address: ]{EOSpace Inc., 6222 185th Avenue Northeast, REDMOND, WA 98052, USA}
\affiliation{National Institute of Standards and Technology, Boulder, CO 80305, USA}
\affiliation{Department of Physics, University of Colorado,	Boulder, CO 80309, USA}

\author{T. R. Tan}
\altaffiliation[Current address: ]{
ARC Centre for Engineered Quantum Systems, School of Physics, The University of Sydney, Sydney, NSW, 2006, Australia
}
\affiliation{National Institute of Standards and Technology,  Boulder, CO 80305, USA}
\affiliation{Department of Physics, University of Colorado,	Boulder, CO 80309, USA}



\author{P.-Y. Hou}
\affiliation{National Institute of Standards and Technology, Boulder, CO 80305, USA}
\affiliation{Department of Physics, University of Colorado,	Boulder, CO 80309, USA}



\author{D. J. Wineland}
\affiliation{National Institute of Standards and Technology, Boulder, CO 80305, USA}
\affiliation{Department of Physics, University of Colorado,	Boulder, CO 80309, USA}
\affiliation{Department of Physics, University of Oregon, Eugene, OR 97403, USA}

\author{A. C. Wilson}
\affiliation{National Institute of Standards and Technology, Boulder, CO 80305, USA}

\author{D. Leibfried}
\affiliation{National Institute of Standards and Technology, Boulder, CO 80305, USA}


\date{\today}

\begin{abstract}
Scaling quantum information processors is a challenging task, requiring manipulation of a large number of qubits with high fidelity and a high degree of connectivity. 
For trapped ions, this could be realized in a two-dimensional array of interconnected traps in which ions are separated, transported and recombined to carry out quantum operations on small subsets of ions.
Here, we use a junction connecting orthogonal linear segments in a two-dimensional (2D) trap array to reorder a two-ion crystal. 
The secular motion of the ions experiences low energy gain and the internal qubit levels maintain coherence during the reordering process, therefore demonstrating a promising method for providing all-to-all connectivity in a large-scale, two- or three-dimensional trapped-ion quantum information processor.
\end{abstract}


\maketitle

\section{Introduction}
Coherent manipulation of trapped atomic ions enables applications ranging from quantum sensing (e.g. force and field sensing, precision spectroscopy, optical clocks) to quantum information processing. 
Most applications must deal with the difficulty in controlling multiple ions, where  for optical clocks the number of clock ions limits the frequency stability \cite{brewer_$^27mathrm^+$_2019}, and for quantum information processing the number of qubits limits the processing capability. 
One approach to increase the number of ions in a system is 
to confine individual ions or small groups of ions in separate trap zones of an array.
Ions are then connected either through probabilistic ion-photon coupling \cite{duan_scalable_2004,monroe_large_2014}, or as discussed here, by employing
the ``quantum charge-coupled device'' (QCCD) architecture \cite{wineland_experimental_1998,kielpinski_architecture_2002}, in which ions are  transported throughout the array to provide the high connectivity required for efficient implementation of general algorithms. 
This approach can extend the features of small ion crystals, such as high-fidelity quantum gates \cite{ballance_high-fidelity_2016,gaebler_high-fidelity_2016} and precise preparation and characterization of the motional state \cite{chen_sympathetic_2017}, to a larger number of qubits.

\begin{figure}[t]
	\centering
	\includegraphics[width=0.45\textwidth]{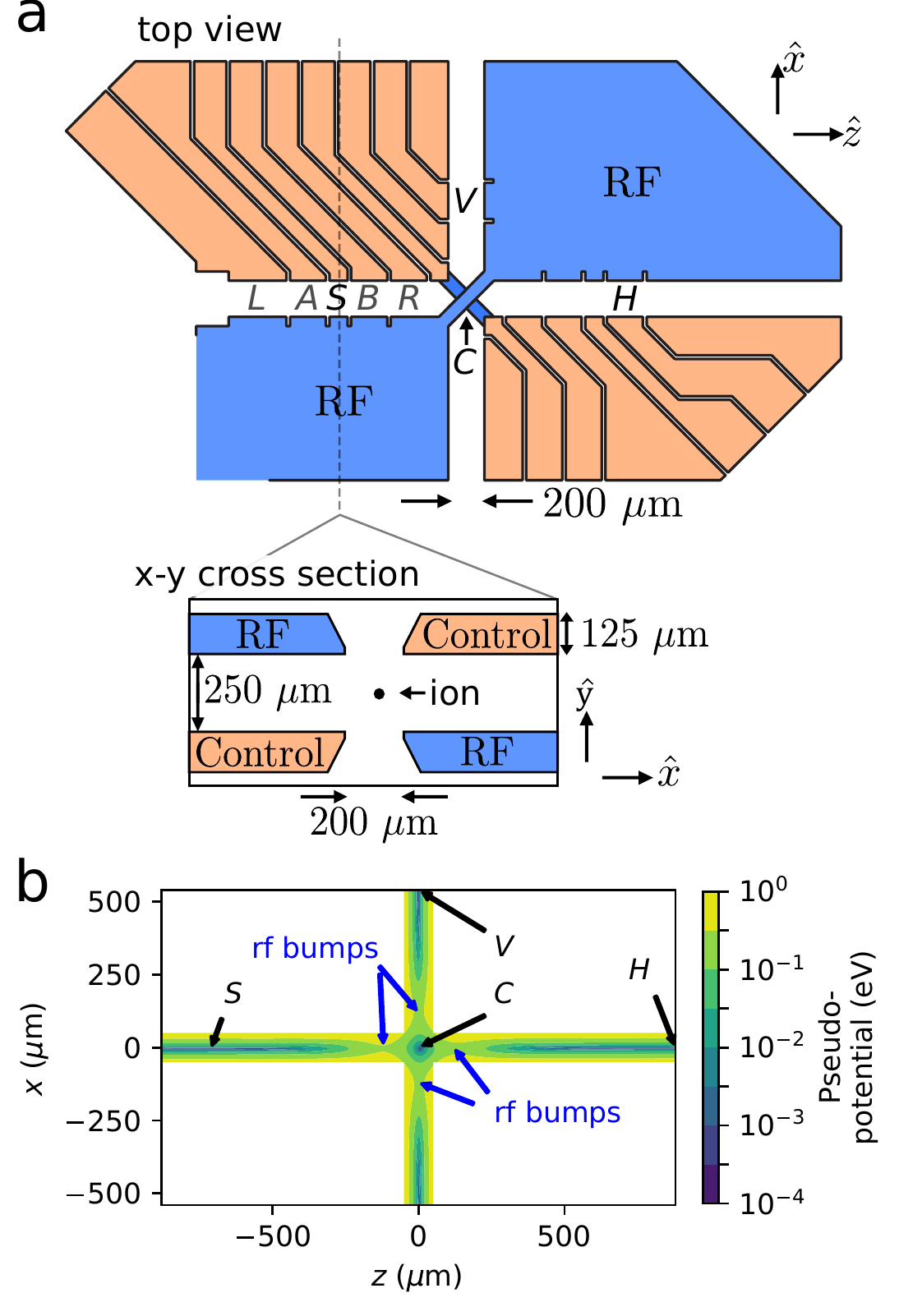}
	\captionmacro[Schematic of the X-junction trap.]{
		{\bf (a)} Schematic view of the top wafer of the trap, with DC (control) electrodes in orange and RF electrodes in blue. A second wafer below the top wafer has DC and RF electrodes swapped, as indicated in the cross-section. 
		The ion shown is located on the axis $x=y=0$ of a linear portion of the trap.
		The ions are held in three major experiment zones, labeled by \szone, \hzone, and \vzone,  connected by the junction located at \czone. Trapping zones $L$, $A$, $B$, and $R$ lying in the same linear region as \szone are used together with the zone \szone to perform operations such as separation, recombination, and individual addressing and detection. (See text for more details.)
		{\bf (b)} Pseudo-potential 
		along the linear channels connected by the junction
		in the plane equidistant to the two wafers defined to be $y=0$. 
		The junction gives rise to four pseudo-potential bumps (indicated by blue arrows) around \czone.
		}  
	\label{fig:schematic_junction_trap}
\end{figure} 

One of the key elements of the QCCD architecture is the ability to reconfigure ion crystals and to hold subsets of ions in different locations, ensuring mutual isolation, while operating on them in parallel. High connectivity and parallelism are considered to be crucial for large-scale fault-tolerant quantum computation \cite{bermudez_assessing_2017}. This requires separation, transport, and rearrangement of ions throughout multiple trapping zones.
Previous experiments have demonstrated adiabatic transport of both single ions \cite{blakestad_high-fidelity_2009} and chains of ions \cite{blakestad_near-ground-state_2011}, diabatic transport and separation \cite{bowler_coherent_2012,walther_controlling_2012}, and fast swapping of neighboring ions in a one-dimensional (1D) array by rotating two ions in place \cite{splatt_deterministic_2009,kaufmann_fast_2017,van_mourik_coherent_2020}. 
These primitives have enabled a transport-based quantum logic gate \cite{leibfried_transport_2007,de_clercq_parallel_2016}, scalable creation of multi-partite entanglement with bipartite interactions \cite{kaufmann_scalable_2017},  quantum-state-assisted sensing \cite{ruster_entanglement-based_2017},  tests of local realism \cite{tan_chained_2017}, and quantum gate teleportation \cite{wan_quantum_2019}. 

Specific to developing a trapped-ion quantum computer, a multi-dimensional trap array is desired to fully realize the power of the QCCD architecture, in which multiple linear trap segments are connected by junctions \cite{hensinger_t-junction_2006,blakestad_high-fidelity_2009,blakestad_near-ground-state_2011,wright_reliable_2013,amini_toward_2010,shu_heating_2014,moehring_design_2011}. Such a multi-dimensional trap array enables smaller average distances between ions than in lower-dimensional architectures and efficiently extends the all-to-all coupling between ions in a small chain \cite{linke_experimental_2017} to connecting arbitrary subsets of a larger number of qubits.

Previous work in traps featuring a junction demonstrated the low-temperature shuttling of ions through RF junctions and characterized the resulting kinetic energy increase \cite{blakestad_high-fidelity_2009,blakestad_near-ground-state_2011}.
Ion crystal reconfigurations were reported in
\cite{hensinger_t-junction_2006,moehring_design_2011} (using junctions but without measurement of kinetic energy changes) or using crystal rotation in a 1D architecture \cite{splatt_deterministic_2009,kaufmann_fast_2017,van_mourik_coherent_2020}.
Here, we use a junction to distribute \be ions in a two-dimensional (2D) architecture and reorder two ions initially in the same potential by combining adiabatic transport and separation primitives. 
We show that the coherence in the internal states of the ions is maintained during the process and characterize excess motional excitation.

\section{Experimental Setup}
\begin{figure}
	\centering
	\includegraphics[width=0.35\textwidth]{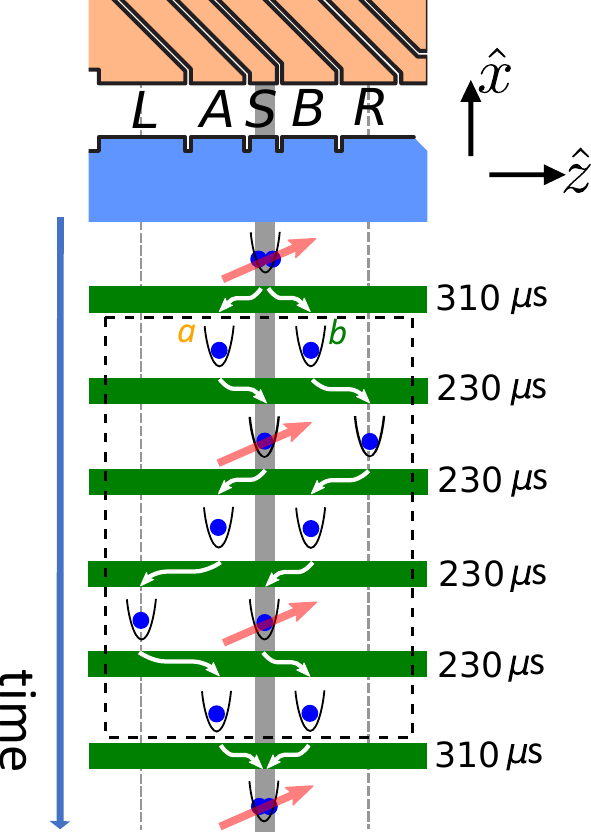}
	\captionmacro[Individual addressing and detection sequence.]{
        Using the sequence depicted within the dashed lines, ions in each well can be manipulated and detected individually, with the shuttling operations in the dashed box taking approximately $1~\mathrm{ms}$. 
        Ion $a$ and $b$ are first separated into zones \ezone and \epzone, and then $a$ is shuttled into \szone while $b$ moves to zone $R$. This formation allows for internal state manipulation and detection of $a$. When $a$ is shifted to $L$, ion $b$ enters \szone and can be manipulated and/or its state detected.
	}  
	\label{fig:separation_shimx}
\end{figure}

A schematic of the trap array is shown in \cref{fig:schematic_junction_trap}a.
The physical structure consists of two electrode wafers separated by $250~\mu$m, providing strong confinement along all three directions, and a third layer ($500~\mu$m below the bottom layer) possessing one single electrode that serves as a common bias electrode across the entire trap.
The trap features three linear regions along the $-z$, $+z$, and $+x$  directions (containing zones \szone, \hzone, and \vzone respectively) with the origin located at the center of the RF junction (\czone) in a plane that is parallel to, and in the middle between the two electrode wafers. 
All other relevant zones ($L$, \ezone, \epzone, $R$) for this work are marked in \cref{fig:separation_shimx}. 
The X-shaped junction (X-junction) at \czone allows us to route ions to \szone, \hzone, and \vzone. Deviation from an ideal infinitely-long linear Paul trap gives rise to pseudo-potential ``bumps'' 
around \czone along the $x$- and $z$-directions
as illustrated in \cref{fig:schematic_junction_trap}b. 
Besides complicating ion transport, this has further implications for quantum information experiments, such as a position-dependent RF modulation index \cite{berkeland_minimization_1998} which affects laser operations and position-dependent qubit frequencies from trap-RF-induced AC-Zeeman shifts. 
These pseudo-potential bumps can also introduce additional motional heating originating from noise in the RF gradients \cite{blakestad_high-fidelity_2009}.
More details on the trap can be found in \cite{blakestad_high-fidelity_2009, blakestad_transport_2010, blakestad_near-ground-state_2011}.

All laser beams for coherent manipulation, state preparation, and state detection are focused to \szone with waists of approximately $25\,\mu$m, while ions in nearby trapping zones are at least $390\,\mu$m away during illumination of ions in \szone. 
We encode qubits in first-order magnetic-field-insensitive hyperfine ground states of \be ions, with $\ket{F=1,m_F=1} \equiv \ket{\uparrow}$ and $\ket{2,0} \equiv \ket{\downarrow}$ \cite{langer_long-lived_2005}. 
Prior to state detection, $\ket{\uparrow}$ is transferred to $\ket{2,2} \equiv \ket{\text{Bright}}$ and $\ket{\downarrow}$ to $\ket{1,-1} \equiv \ket{\text{Dark}}$. 
A resonant laser driving the $S_{1/2} \ket{2,2} \leftrightarrow P_{3/2} \ket{3,3}$ cycling transition is used to distinguish the two states through photon counts \cite{nagourney_shelved_1986,sauter_observation_1986,bergquist_observation_1986}.

The motional heating and excess energy accumulated during certain transport primitives 
is investigated by running the test sequences listed in \cref{tab:summary-waveforms}. 
To characterize a primitive, a single \be ion or two \be ions are initialized in \szone by cooling the axial modes ($\omega_\mathrm{COM} = 2\pi\times 3.6$ MHz, $\omega_\mathrm{STR} = 2\pi\times 6.2$ MHz) of the ions close to the motional ground state 
($\bar{n}=0.016(2)$ for a single ion, 
$\bar{n}_\mathrm{COM}=0.038(9)$ and $\bar{n}_\mathrm{STR}=0.014(7)$ for two ions).
The radial modes (mode frequencies $11$ -- $13$ MHz) are left close to the Doppler temperature corresponding to an average motional occupation number $\bar{n}$ of approximately $0.5$. 
After completing a transport test sequence and returning to \szone, the final state of the ion motion is probed with motion-sensitive Raman transition beams on blue and red sidebands \cite{meekhof_generation_1996,meekhof_generation_1996b} in separate experiments. 
Assuming a thermal distribution of final energies, we extract $\bar{n}$ of the ion motional modes by fitting a Rabi-oscillation model to the data from both experiments. 
More details of the motional state analysis are provided in the supplementary materials.

We now describe several of the experimentally implemented transport primitives in more detail (see also Table \ref{tab:summary-waveforms} for a summary). 
We trap two \be ions ($a$ and $b$) in a single well \szone and subsequently separate them from an initial spacing of about $5\,\mu\mathrm{m}$ into two individual wells located at \ezone and \epzone.
The well minima are separated by $\sim 340\,\mu$m and are formed in $\sim 310\,\mu$s by ramping the harmonic and quartic terms of the potential \cite{home_electrode_2006}.
A suitable static electric field along the axial direction of the crystal, superimposed on the separation waveform, shifts the center of the ion crystal relative to the center of the quartic potential, enabling control over the number of ions transported into the respective individual wells \cite{bowler_coherent_2012}. 
When separating two ions into wells \ezone and \epzone, this primitive is denoted as $\Sab \ra \AaBb$ in Table \ref{tab:summary-waveforms} (row 6). Here, the symbols before and after the right arrow denote the initial and final configuration of the primitive, respectively. The capital letters of a configuration denote the positions of the wells, and the subscripts denote the ion(s) residing in each well, respectively.
Each ion travels a distance of approximately $167.5~\mu\mathrm{m}$. To characterize $\Sab \ra \AaBb$, we run a longer test sequence $\Sab \ra \AaBb \ra \Sab$, which is implemented by concatenating the forward and reversed version of the primitive $\Sab \ra \AaBb$, and measure the excitation in the center-of-mass (COM) and stretch (STR) mode after recombination of $a$ and $b$ in \szone
(the well in which the motional state is characterized at the end of a test sequence is underlined in Table \ref{tab:summary-waveforms}). 
We find an average COM mode occupation of $\bar{n}$=0.55(3) and 0.43(3) in the STR mode.

An important shuttling sequence, illustrated in the dashed box in \cref{fig:separation_shimx}, will be referred to as the ``individual addressing and detection sequence'':
\begin{equation}
\label{eqn:detection_sequence}
\AiBj \ra \SiRj \ra \AiBj \ra \LiSj \ra \AiBj\,.
\end{equation}
Additional laser pulses can be applied during configurations \SiRj and \LiSj to manipulate the ions' internal states individually after separation and before detection (\cref{fig:separation_shimx}). This allows us to individually rotate each ion on its qubit Bloch sphere.
The total duration of the transport in the individual addressing and detection sequence, not including any manipulation or detection operations, is about $1$~ms. This sequence can also be used to determine the number of ions in each well.
       
Combining $\Sab \ra \AaBb$ with 
the individual addressing and detection sequence
allows us to probe the temperature of individual ions. 
To this end, motion-sensitive Raman beams are applied in the configuration \SaRb to determine the temperature of the ion $a$ in the left well, and \LaSb for the ion $b$ in the right well, deriving average occupation numbers of 0.10(1) and 0.25(2), respectively, using sideband thermometry \cite{meekhof_generation_1996}. 

\begin{table*}
	\centering
	\captionmacro[Summary of transport primitives for \be ions.]{To characterize the performance of each primitive, a test sequence is used where the motional excitation is measured in one well configuration (underlined) 
		towards the end or in the middle of the test sequence. At the beginning of the test sequence, the axial modes of ion crystals are cooled close to the motional ground states ($\bar{n}=0.016(2)$ for a single ion, $\bar{n}_\mathrm{COM}=0.038(9)$ and $\bar{n}_\mathrm{STR}=0.014(7)$ for two ions).  
		The symbols \czone and \cyzone here indicate the same trapping zone, but with the weakest axis of the trapping potential aligned with the $z$-axis and $x$-axis, respectively. 
		Subtracting the initial $\bar{n}$ after ion preparation and excitations during common sections in the test sequences allows us to derive the excess motional excitation per transport primitive $\Delta n_p$. If a primitive is run forward and backward, we assume that these two parts contribute equally to $\Delta n_p$. The test sequence for determining the motional excitation of the primitive $\Sab \ra \AaBb$ (row 6) includes mode mixing in the process of separation and recombination, therefore no values for $\Delta n_p$ are derived. As a consequence, there is no prediction of motional excitations in the well configuration \AaBb, and the measurement results in row 7-8 are used as the baseline for the test sequences in the rows below.
	}
	\begin{tabular}{c l l c c c c c c}  
		\toprule
		 \# & Primitive & Test Sequence & Crystal & $\bar{n}$ & Duration ($\mu$s) & Distance ($\mu$m) & $\Delta n_\mathrm{p}$\\
		\midrule
		1& \Sa \ra \Aa & \Sa \ra \Aa \ra \Sua & \be & 0.045(3) & 68 & 170 & 0.015(2) \\
		2& \Aa \ra \Ca & \Sa \ra \Aa \ra \Ca \ra \Aa \ra \Sua & \be & 0.12(1) & 128 & 880 & 0.038(5) \\
		\multirow{2}{*}{3}&\multirow{2}{*}{\Ha \ra \Ca} & \multirow{1}{*}{\Sa \ra \Aa \ra \Ca } & \multirow{2}{*}{\be} & \multirow{2}{*}{0.27(1)} & \multirow{2}{*}{128} & \multirow{2}{*}{880} & \multirow{2}{*}{0.075(7)} \\
		&& \multicolumn{1}{r}{\ra \Ha \ra \Ca \ra \Aa \ra \Sua}  &&&&&\\
		\multirow{2}{*}{4} & \multirow{2}{*}{\Ca \ra \Cya} & \Sa \ra \Aa \ra \Ca & \multirow{2}{*}{\be} & \multirow{2}{*}{0.23(1)} & \multirow{2}{*}{57} & 		\multirow{2}{*}{-} & \multirow{2}{*}{0.055(7)} \\
		&&\multicolumn{1}{r}{\ra  \Cya \ra \Ca \ra \Aa \ra \Sua}&&&&&\\ 
		\multirow{2}{*}{5}&\multirow{2}{*}{\Va \ra \Ca} & \Sa \ra \Aa \ra \Ca \ra \Cya \ra \Va & \multirow{2}{*}{\be} &
		\multirow{2}{*}{0.30(2)} & \multirow{2}{*}{132} & \multirow{2}{*}{540} & \multirow{2}{*}{0.04(1)} \\
		&&\multicolumn{1}{r}{\ra \Cya \ra \Ca \ra \Aa \ra \Sua}&&&&& \\
		\multirow{2}{*}{6}&\multirow{2}{*}{\Sab \ra \AaBb} & \multirow{2}{*}{\Sab \ra \AaBb \ra \Suab} & \multirow{2}{*}{\be-\be} & 0.55(3) (COM) &
		\multirow{2}{*}{310} & \multirow{2}{*}{167.5/167.5} & \multirow{2}{*}{-} \\
		&&&&0.43(3) (STR)&&  & \\
		\multirow{2}{*}{7}&\multirow{2}{*}{\AaBb \ra \SaRb} & \Sab \ra \AaBb \ra \SuaRb   & \multirow{2}{*}{\be-\be} & \multirow{2}{*}{0.10(1)} & \multirow{2}{*}{230} & \multirow{2}{*}{160/220} & \multirow{2}{*}{-}
		\\
		&&\multicolumn{1}{r}{\ra \AaBb \ra \LaSb \ra \AaBb \ra \Sab}&&&&&\\
		\multirow{2}{*}{8}&\multirow{2}{*}{\AaBb \ra \LaSb} & \Sab \ra \AaBb \ra \SaRb   & \multirow{2}{*}{\be-\be} & \multirow{2}{*}{0.25(2)} & \multirow{2}{*}{230} &
		\multirow{2}{*}{280/160} & \multirow{2}{*}{-} \\
		&&\multicolumn{1}{r}{\ra \AaBb \ra \LaSub \ra \AaBb \ra \Sab}&&&&&\\
		\multirow{2}{*}{9} & \multirow{2}{*}{\AaBb \ra $A_aC_b$} & \Sab \ra \AaBb \ra $A_aC_b$ \ra \AaBb \ra \SuaRb& \multirow{2}{*}{\be-\be} &  \multirow{2}{*}{0.13(1)} & \multirow{2}{*}{110} & \multirow{2}{*}{0/540} & \multirow{2}{*}{0.015(7)} \\ 
		&&\multicolumn{1}{r}{\ra \AaBb \ra \LaSb \ra \AaBb \ra \Sab}& &&&& \\ 
		\multirow{2}{*}{10} & \multirow{2}{*}{\AaBb \ra $A_aC_b$} & \Sab \ra \AaBb \ra $A_aC_b$ \ra \AaBb \ra \SaRb & \multirow{2}{*}{\be-\be} &  \multirow{2}{*}{0.63(4)} & \multirow{2}{*}{110} & \multirow{2}{*}{0/540}&  \multirow{2}{*}{0.19(2)} \\ 	
		&&\multicolumn{1}{r}{\ra \AaBb \ra \LaSub \ra \AaBb \ra \Sab}&&&&& \\ 
		\bottomrule
	\end{tabular}
	\label{tab:summary-waveforms}
\end{table*}

\begin{figure*}
	\centering
	\includegraphics[width=0.95\textwidth]{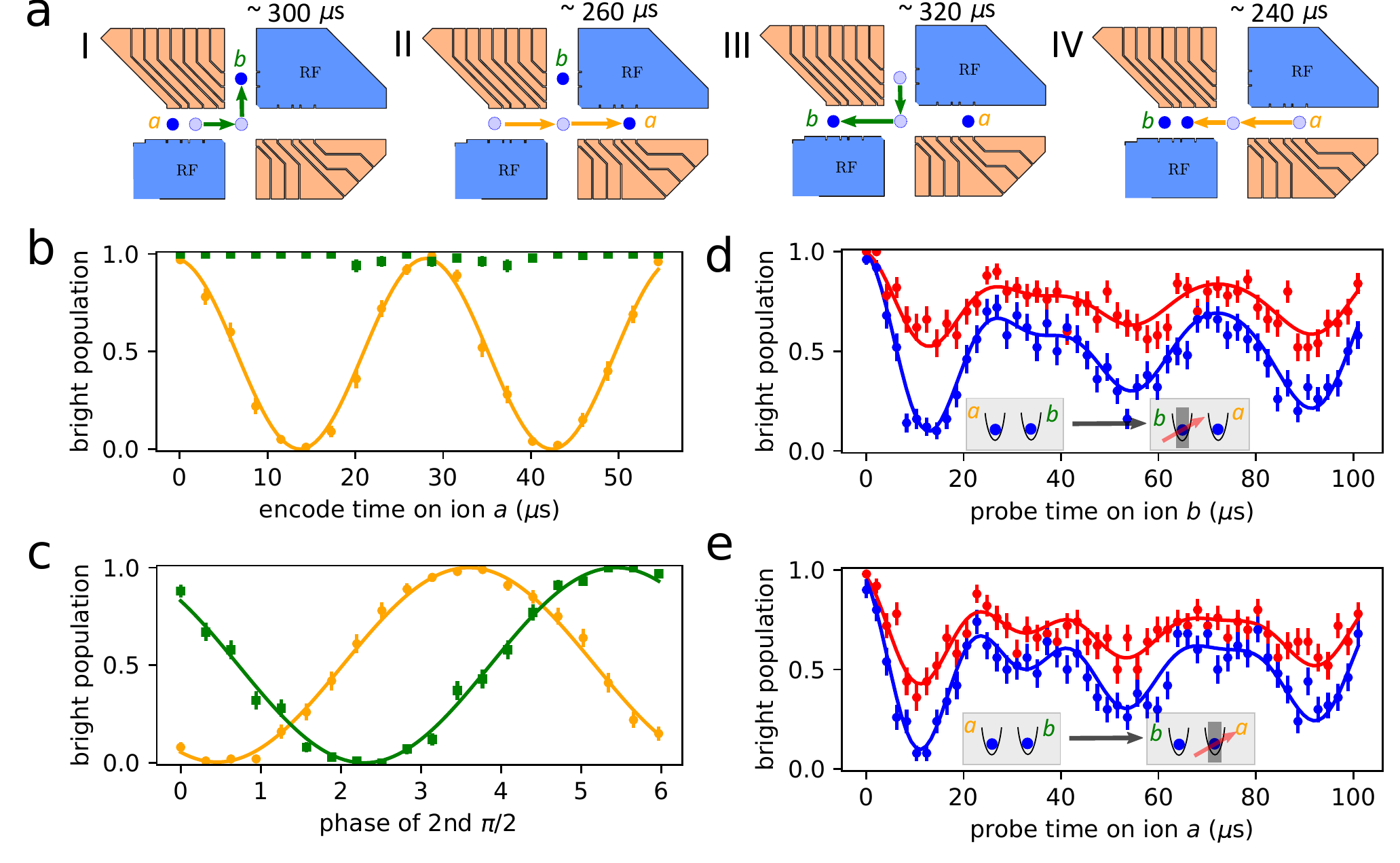}
    	\captionmacro[Reordering two \be ions using the X-junction.]{
    	{\bf (a)} Schematic representation of reordering sequence. Two ions $a$ and $b$ in the double well potential are shuttled sequentially through the junction to separated regions of the trap array, and then moved back to the initial well with their order swapped.
		The arrows (orange for ion $a$ and green for ion $b$) indicate the trajectories of each ion (light blue circles) and the blue circles represent the end points of the primitives on the trajectories.
		{\bf (b)} Ion $a$ is excited using a single-qubit rotation in the configuration \SaRb, and a Rabi oscillation is observed in $S_aL_b$ (orange) after the reordering sequence. No population oscillation is observed for detection performed in $R_aS_b$ (green), while ion $b$ is ideally in \szone. 
		{\bf (c)}
		Coherence of the internal states of ion $a$ (orange) and $b$ (green) is maintained in two corresponding Ramsey sequences enclosing an exchange  of ion positions and addressing one of the ions respectively. The phase shifts ($0.46(2)$ rad for ion $a$ (orange) as the offset of the fringe minimum from $0$ and $2.29(2)$ rad for ion $b$ (green)) arise mainly from the durations that the two ions accumulate phase due to a frequency shift relative to the local oscillator. See text for more details. 
		{\bf (d-e)}
		The temperatures of the two \be ions are probed on the red (red dots) and blue (blue dots) sidebands  after the reordering sequence.
		Fits to the Rabi oscillation model outlined in the supplementary material (solid lines) result in average occupation numbers of 1.1(1) for ion $b$ and 1.7(1) for ion $a$.
	} 
	\label{fig:reorder-flop}  
\end{figure*} 

\section{reordering two ions}
By combining further transport primitives (see Table \ref{tab:summary-waveforms} for their individual characteristics), we demonstrate reordering of a two-ion crystal by separating ions $a$ and $b$ and then moving the ions around each other with the aid of the X-junction. 
This is done by separating a two-ion crystal $\Sab \ra \AaBb$, shuttling of ion $b$ to \vzone ($\AaBb \ra A_aC_b \ra A_aV_b$, step I in \cref{fig:reorder-flop}a), shuttling ion $a$ to \hzone ($A_aV_b \ra C_aV_b \ra H_aV_b$, step II), moving ion $b$ to \ezone ($H_aV_b \ra H_aC_b \ra H_aA_b$, step III), moving ion $a$ to $B$ ($H_aA_b \ra C_aA_b \ra \BaAb$, step IV), and combining $a$ and $b$ ($\BaAb \ra \Sba$).
The full reordering sequence reads
\begin{align}
\begin{split}
\label{eq:reorder-sequence}
\Sab \ra \AaBb \ra 
A_aC_b \ra A_aV_b \ra C_aV_b \ra H_aV_b \\
 \rightarrow H_aC_b \rightarrow H_aA_b \rightarrow C_aA_b  \rightarrow \BaAb \rightarrow \Sba  
\end{split}
\end{align}
with the duration of each individual segment listed in \cref{tab:summary-waveforms}. 
The duration of the reordering sequence without separation and recombination (\AaBb to \BaAb) is about $1.1$~ms. 

The transport waveforms used in this paper require moving one potential well while holding the second potential well stationary.
To avoid cross talk between different wells, the potential governing all trapping zones is considered when generating the waveforms. 


We can insert the individual addressing and detection sequence at \AaBb or \BaAb in sequence (\ref{eq:reorder-sequence}) to encode the spin state of each ion before reordering and to detect the ion positions after reordering. The encoding is performed using a pair of co-linear Raman beams in the configurations \SaRb and \LaSb, addressing only the ions located at \szone. 
To verify a position swap between the two ions, we first apply a laser pulse for various durations to Rabi-flop ion $a$ in the configuration \SaRb. 
After the reordering sequence, a population oscillation as a function of pulse duration is detected only in $S_aL_b$ (on ion $a$), and not in $R_aS_b$ (on ion $b$) as shown in \cref{fig:reorder-flop}b, indicating  successful reordering. 
Rabi-flopping ion $b$ in the same fashion yields a similar result, now with $b$ exhibiting population oscillations and $a$ remaining in the same state.

We also show that qubit coherence is maintained after reordering by applying a Ramsey sequence to each ion individually. We apply a $\pi/2$-pulse on ion $a$ ($b$) in the configuration \SaRb (\LaSb), execute the reordering, and apply the second $\pi/2$-pulse with a variable phase relative to that of the first $\pi/2$-pulse by addressing ion $a$ ($b$) in the reverse configuration $S_aL_b$ ($R_aS_b$).
The resulting Ramsey fringes show a contrast close to $1$ for both ions addressed as shown in \cref{fig:reorder-flop}c.
The phase shift is mainly caused by the different durations that the two ions integrate over a $-1.7~\mathrm{kHz}$ frequency difference between the qubits and the local oscillator (corresponding to phase shifts of $\approx -18.2$ rad for ion $a$ and $\approx -28.7$ rad for ion $b$). 
The frequency difference is due to the AC-Stark shift on the qubit transition from the laser beams implementing the Ramsey pulses, whose frequency difference is set to be on resonance while rotating the state of the ions in the presence of the AC-Stark shift.
In addition, AC-Zeeman shifts and a magnetic field gradient across the trap are non-negligible when shuttling ions across millimeter length scales in our setup. 
A separate investigation shows that the AC-Zeeman shifts on the qubit transition $\left(\ket{\uparrow}\leftrightarrow\ket{\downarrow}\right)$ of \be vary by $10~\mathrm{Hz}$ over a distance of about $15~\mu\mathrm{m}$ along the $x$-direction at \szone.
From a measured static magnetic field gradient of approximately $4.1\times10^{-3}~\mathrm{T/m}$ and the second-order field sensitivity coefficient $c_2 = 3.05\times10^{-1}~\mathrm{Hz/\mu T^2}$ \cite{langer_long-lived_2005}, we estimate a qubit frequency shift of $\approx 1~\mathrm{mHz}$ over the same distance of $15~\mathrm{\mu m}$ \cite{langer_long-lived_2005,langer_high_2006}. 
These effects will need to be minimized, properly calibrated, or reduced by dynamical decoupling or error correction in future large-scale devices. 

The axial temperature of each ion after reordering, measured through sideband thermometry,  indicates that the full sequence introduces an average motional excitation of 1.1(1) for ion $b$ and 1.7(1) for ion $a$ (\cref{fig:reorder-flop} d-e). 
The measured motional excitation after reordering is a factor of 2-3 times larger than the values obtained by summing up the excitation of the constituent primitives. 
We believe that the additional excitation can be attributed to 
non-continuous concatenation and heating during idle periods in static wells, but this requires further experimental and theoretical study.

\begin{figure*}
	\centering
	\includegraphics[width=0.80\textwidth]{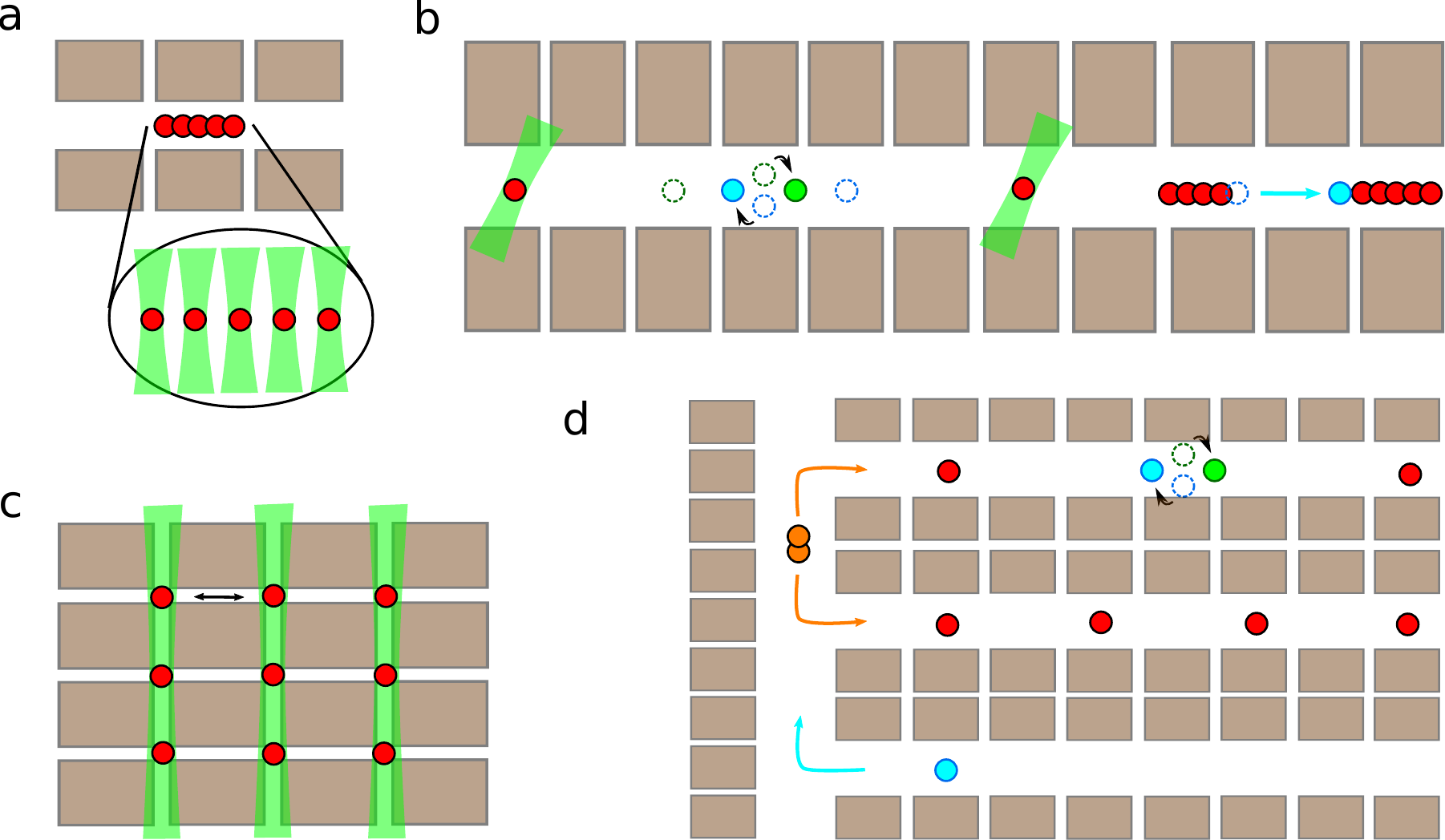}
    	\captionmacro[Chip-based multi-qubit quantum devices.]{
    	{\bf (a)} 1D chain of individually-addressed ions. Tightly focused beams allow individual addressing of ions in the chain.
		{\bf (b)} 1D trap array. Ions are confined in separated wells created by electrode arrays. Separation between the wells that is large compared to the electrode dimensions allows for isolated control of each well. Coupling of a certain pair of ions are realized by repeated swap operations or crystal rotation (black arrows) and separation/recombination. 
		{\bf (c)} Multi-dimensional trap lattices. Ions are confined in fixed potential wells, while the coupling between the ions is enabled by tuning the potentials (black arrow).
		{\bf (d)} Multi-dimensional trap array. Couplings between the ions are realized by shuttling information carriers (blue) or messengers (orange) through dedicated sections of the array. 
		} 
	\label{fig:architecture}  
\end{figure*} 

\section{Potential large-scale configurations}
Here, we discuss several architectures for building multi-qubit quantum devices using trapped ions. Related architectures are considered in \cite{ratcliffe_scaling_2018,welzel_designing_2011}.

{\bf a) 1D chain of individually-addressed ions (\cref{fig:architecture}a).}  All ions are trapped in a single well and are connected by the long-range Coulomb interaction, while  individual control of each ion is realized by tightly-focused laser beams. A theoretical study determined the method to be applicable for an arbitrarily long chain of ions \cite{landsman_two-qubit_2019}, and experimentally, two-qubit gates between any two qubits within a 11-qubit ion chain  have been performed \cite{wright_benchmarking_2019}. 

{\bf b) 1D trap array (\cref{fig:architecture}b).} As an extension of the 1D chain in a single well, one can confine the ions within a 1D array of potential  wells and apply global rotations on ions in each single well. The ions in different wells are then connected via linear shuttling and separation combined with swap gates or crystal  rotations \cite{kaufmann_fast_2017}. 
In such a 1D trap array, it takes $O(n)$ such rotations to transport an ion across $n$ other ions using a sequence of crystal rotations. 
Considering the equivalence of re-ordering and swap operation, this will induce an overhead (longer circuit depth) on the order of $O(n)$ when rearranging a quantum circuit of $n$ qubits with all-to-all connectivity to a corresponding circuit with only linear nearest neighbor interactions \cite{translation_cheung_2007}.
By confining multiple ions in each segment, one can combine the all-to-all coupling as in architecture a) and the coupling between segments through shuttling. This method is predicted to achieve fault tolerance with  modest  reduction  of  the  fault  tolerance  threshold compared to the 2D surface code despite the topological restriction posed by the 1D architecture \cite{li_one-dimensional_2018}.

{\bf c) Multi-dimensional trap lattices (\cref{fig:architecture}c).} Individual trapping zones are held in fixed positions in space, forming a lattice of singly-occupied confining potentials with nearest-neighbor couplings tuned by bringing the traps in and out of resonance. The higher dimension provides each qubit with a larger number of nearest neighbors  \cite{schmied_optimal_2009,krauth_optimal_2014}. 
An open electrode geometry, as for example provided by surface electrode traps \cite{chiaverini_surface-electrode_2005} is advantageous to not substantially reduce nearest neighbor coupling by shielding from nearby electrodes.
This increased connectivity between the ions therefore reduces the overhead in circuit depth for implementing a quantum circuit 
\cite{translation_cheung_2007}. For example, employing a two-dimensional square lattice where each ion has four nearest neighbors already reduces this overhead to $O(\sqrt{n})$, while 3D square lattice shares a similar feature with a reduced overhead of $O(\sqrt[3]{n})$ \cite{translation_cheung_2007}.

{\bf d) Multi-dimensional trap array (\cref{fig:architecture}d).}  A multi-dimensional trap array could combine all modes of operation discussed in a)-c) with all-to-all connectivity between the ions by shuttling through dedicated information highways, while entangled pairs of resource ancillas can be distributed ahead of time for quantum gate teleportation to reduce the latency required by the shuttling process itself \cite{wan_quantum_2019}. 
In comparison to the 1D architecture, the geometry of the trap  in such an multi-dimensional architecture also reduces the average distance between any pair of ions.
 Small-scale  reconfiguration  within  linear  regions  of   the multi-dimensional array would most efficiently be done through swap gates or crystal rotation as discussed in b), while connecting distant qubits or replacing lost ions would involve moving through junctions and dedicated transport highways.
Such an architecture more capable in reconfiguration of ion crystals will likely be required for the construction of multiple logical qubits \cite{bermudez_assessing_2017}.

\section{Summary}
By combining separation and shuttling primitives, we were able to reorder two \be ions using an RF junction connecting three trap zones on different sides of that junction.
We verified the reordering using transport-assisted individual addressing, and showed that quantum coherence encoded on individual qubits was maintained during the reconfiguration. We also briefly discussed example configurations for large-scale devices.

\begin{acknowledgments}
We thank P.~Kent and J.~F.~Niedermeyer of NIST for helpful comments on the manuscript.
This work was supported by the Office of the Director of National Intelligence (ODNI) Intelligence Advanced Research Projects Activity (IARPA), ONR, and the NIST Quantum Information Program.
S.D.E.\ acknowledges support by the U.S. National Science Foundation under Grant No. DGE 1650115.
Y.W., S.D.E., and J.J.W. are associates in the Professional Research Experience Program (PREP) operated jointly by NIST and University of Colorado Boulder. 
The authors declare that there are no competing financial interests.
\end{acknowledgments}



%


\clearpage
\section{Supplementary Materials}

\subsection{Waveform generation}
We use the package {\it bem} \cite{bem_2019} that uses the boundary element method  to generate a potential map for each individual electrode, while all other electrodes are held at ground potential. This potential map includes both the effect of electrostatic potentials from DC electrodes and the contribution of pseudo-potentials from RF electrodes.
Starting from this map, we construct the potentials experienced by the ions while several electrodes are held at non-zero potential.
Vanishing first derivatives of the potentials define the saddle points and equilibrium positions for single ions, positive second derivatives define  trap frequencies, and third and fourth derivatives need to be taken into account during ion separation. 
We use the package {\it electrode} \cite{electrode_2019} (which uses {\it cvxopt} \cite{cvxopt_2019}) to produce waveforms (a set of potentials, with each member applied to a certain electrode as a function of time) to fulfill the full set of constraints that define a transport primitive. 
The waveform simulation does not include the Coulomb interaction between ions, so the response of multi-ion crystals is calculated after generating the waveform by performing a normal mode calculation \cite{james_quantum_1998}. 
Before applying the waveforms to the trap electrodes, we pre-compensate the waveforms \cite{bowler_arbitrary_2013,bowler_coherent_2015} to account for filter distortion from the low-pass filters used to reduce technical noise on the trap electrodes. These waveforms with pre-compensation are generated with an arbitrary waveform generator with 50 MHz update rate \cite{bowler_arbitrary_2013}.

\subsection{Sideband thermometry}
Ion crystal temperature is determined by probing the red- and blue-sideband (RSB and BSB) transitions followed by a state-dependent fluorescence detection. 
For a single ion probed along one particular mode on the 1st BSB and 1st RSB, the measurement results are fitted with model functions for the population $P$ remaining in the initial state after driving the sideband for a duration $t$ \cite{meekhof_generation_1996,meekhof_generation_1996b}
\begin{align}
\begin{split}
	P_\mathrm{BSB}(t) & =\sum_{n=0}^{N-1} p_n\cdot \frac{1}{2}\left[1+\cos(\Omega_{n,+1}\, t)  \exp(-\gamma t)\right]\\
	P_\mathrm{RSB}(t) & = p_\mathrm{0} + \sum_{n=1}^{N} p_n\cdot \frac{1}{2}\left[1+\cos(\Omega_{n,-1}\, t)  \exp(-\gamma t)\right]\,,
	\label{eqn:flopping_one_ion}
\end{split}
\end{align}
where $p_n = \frac{\bar{n}^n}{(\bar{n}+1)^{n+1}}$ is the population in each motional number state $\ket{n}$ assuming a thermal distribution with an average phonon number $\bar{n}$. $\Omega_{n,\kappa}$
is the Rabi frequency for $\ket{n}$ on the probed sideband $\kappa$ with $\kappa=+1$ for the first blue sideband and $\kappa=-1$ for the first red sideband, $\gamma$ is a phenomenological decay constant, and $N$ is the quantum number of the highest Fock state $\ket{N}$ considered in the model. 
The Rabi frequency $\Omega_{n,\kappa}$ takes the form
\begin{align}
\begin{split}
    \Omega_{n,\kappa} & = \Omega\, D_{n,\kappa, \eta} \\
    & = \Omega\, \exp{(-\frac{\eta^2}{2})}\left(\frac{n_{<}!}{n_{>}!}\right)^{1/2}\eta^{|\kappa|} L_{n_{<}}^{|\kappa|}(\eta^2)\,,
\end{split}
\end{align}
where $n_{<}$ and $n_{>}$ are the lesser and greater of $n$ and $n+\kappa$, $\eta$ is the Lamb-Dicke parameter, and $L_x^y(z)$ is a generalized Laguerre polynomial.
The decay constant $\gamma$ is introduced as a simplified description of  the decay of Rabi oscillations coming from various effects, including motional dephasing, spontaneous emission, amplitude and phase noise of laser beams, and anomalous heating.
We use this model to extract the average phonon number $\bar{n}$ after initial sideband cooling and optionally after transport of a single ion as shown in \cref{fig:reorder-flop}d-e. 

If more than one mode is involved in the probe process, meaning that Lamb-Dicke parameters for the probe laser beam are non-zero for more than one mode, one needs to include the Debye-Waller-type coupling induced by the spectator modes \cite{wineland_experimental_1998,wan_efficient_2015}.
This is the case, e.g., for a mixed-species two-ion crystal probed along the crystal axis, or for a single ion probed along a direction that has overlap with more than one normal mode direction of the confining potential. 
We model the Rabi oscillations driven on a single ion, but involving two motional modes as
\begin{align}
\begin{split}
	P^\mathrm{(1)}_\mathrm{BSB}(t) & = \sum_{m=0}^{M} \sum_{n=0}^{N-1} p_{nm}\cdot \frac{1}{2}\big[1 \\ 
	& +\cos(\Omega_{nm,+1\,0}\, t)  \exp(-\gamma t)\big] \\
	P^\mathrm{(1)}_\mathrm{RSB}(t) & = \sum_{m=0}^{M} \Big\{ p_{0m} \\
		 & + \sum_{n=1}^{N} p_{nm}\cdot \frac{1}{2}\big[1
		 +\cos(\Omega_{nm,-1\,0}\, t)  \exp(-\gamma t)\big]\Big\}\\
	P^\mathrm{(2)}_\mathrm{BSB}(t) & = \sum_{n=0}^{N} \sum_{m=0}^{M-1} p_{nm}\cdot \frac{1}{2}\big[1 \\ 
	& +\cos(\Omega_{nm,0\,+1}\, t)  \exp(-\gamma t)\big] \\
	P^\mathrm{(2)}_\mathrm{RSB}(t) & = \sum_{n=0}^{N} \Big\{ p_{n0} \\
		 & + \sum_{m=1}^{M} p_{nm}\cdot \frac{1}{2}\big[1
		 +\cos(\Omega_{nm,0\,-1}\, t)  \exp(-\gamma t)\big]\Big\}\,.
	\label{eqn:flopping_two_ions_mix}
\end{split}
\end{align}
The four equations above return the population that remained in the initial state when probed on BSB of 1st mode, RSB of 1st mode, BSB of 2nd mode, and RSB of 2nd mode, respectively, for a duration $t$. 
Here, $p_{nm}$ is the population in the Fock state $\ket{n,m}$, and 
\begin{equation}
    \Omega_{nm,\kappa\lambda} = \Omega\, D_{n,\kappa,\eta_{1}} D_{m, \lambda, \eta_{2}}
\end{equation}
is the Rabi rate of driving a transition $\ket{n, m}\ra \ket{n+\kappa, m+\lambda}$ 
with $\kappa$ and $\lambda$ indicating changes of motional quantum numbers in the two modes, and $\eta_{1}$, $\eta_{2}$ are the Lamb-Dicke parameters of the two modes \cite{wineland_experimental_1998}. The curve fitting is performed on four data curves (first red and blue sidebands on two modes) simultaneously assuming a common set of fitting parameters. 
In particular, this means that even for a mode with $\bar{n}=0$ Rabi oscillations on the BSB will show a decay originating from the statistical distribution in motional states of the other mode. 
Evaluating the Rabi oscillation data using a common set of parameters allows for a more precise determination of the average phonon number 
at low $\bar{n}$ by considering a more complete model and also at high $\bar{n}$
even if
red and blue sideband excitations have only small differences from the bounded decay of the companion mode.


To determine the temperature of two-ion crystals of the same species, we use a model describing the Rabi oscillations of two trapped ions that are simultaneously driven. 
Here, the Rabi oscillations of two ions are considered as the summation of Rabi oscillations of many three-level systems. 
An analytical solution for the problem describing the Rabi oscillations of a three-level system has been provided in \cite{fujii_rabi_2006}. 
For ions fluorescing in $\ket{\uparrow}$ and not fluorescing in $\ket{\downarrow}$, the average fluorescence at a probe duration $t$, normalized to two ions fluorescing in $\ket{\uparrow\uparrow}$, is
\begin{align}
	\overline{P}(t) 
	=  1 \cdot P_{\ket{\uparrow\uparrow}} + \frac{1}{2} \cdot
	P_{\ket{\uparrow\downarrow}+\ket{\downarrow\uparrow}}\,, 
	\label{eqn:flopping_two_ions_same}
\end{align}
and takes following the forms when probed on the BSB of the center-of-mass (COM) mode, RSB of the COM mode, 
the BSB of the stretch (STR) mode, and RSB of the STR mode respectively
\begin{align}
    \begin{split}
       \overline{P}^\mathrm{COM}_\mathrm{BSB}(t) & = 
      \sum_{m=0}^{M} \sum_{n=0}^{N-1} p_{nm}\cdot \big(|c^{(0)}_{nm,+1\,0}|^2 \\
     & +  \frac{1}{2}\cdot|c^{(1)}_{nm,+1\,0}|^2 
     \big) \\
     \overline{P}^\mathrm{COM}_\mathrm{RSB}(t) & = 
      \sum_{m=0}^{M} \Big\{ p_{0m} + \sum_{n=1}^{N} p_{nm} \cdot \big(|c^{(0)}_{nm,-1\,0}|^2 \\
     & +  \frac{1}{2}\cdot|c^{(1)}_{nm,-1\,0}|^2 
     \Big\}
     \\
            \overline{P}^\mathrm{STR}_\mathrm{BSB}(t) & = 
      \sum_{n=0}^{N} \sum_{m=0}^{M-1} p_{nm}\cdot \big(|c^{(0)}_{nm,0\,+1}|^2 \\
     & +  \frac{1}{2}\cdot|c^{(1)}_{nm,0\,+1}|^2 
     \big) \\
     \overline{P}^\mathrm{STR}_\mathrm{RSB}(t) & = 
      \sum_{n=0}^{N} \Big\{ p_{n0} + \sum_{m=1}^{M} p_{nm} \cdot \big(|c^{(0)}_{nm,0\,-1}|^2 \\
     & +  \frac{1}{2}\cdot|c^{(1)}_{nm,0\,-1}|^2 
     \big) \Big\}
     \,,
    \end{split}
\end{align}
where $|c^{(0)}_{nm,\kappa\lambda}|^2$, $|c^{(1)}_{nm,\kappa\lambda}|^2$
are the coefficients describing the evolution of the three-level system. 
Starting from $\ket{\uparrow\uparrow}$, these coefficients take the form
\begin{align}
\begin{split}
	c^{(0)}_{nm,\kappa\lambda} = & \frac{g_{1,nm,\kappa\lambda}^{2} \cos \left(g_{\mathrm{c},nm,\kappa\lambda} t\right) +g_{2,nm,\kappa\lambda}^{2}}{g_{\mathrm{c},nm,\kappa\lambda}^2}\\
	c^{(1)}_{nm,\kappa\lambda} = & -i \frac{g_{1,nm,\kappa\lambda} \sin \left(g_{\mathrm{c},nm,\kappa\lambda} t\right) }{g_{\mathrm{c},nm,\kappa\lambda}}
\end{split}
\end{align}
according to  Eq.~10 in \cite{fujii_rabi_2006} up to a phase factor (see Eq.~11 in \cite{fujii_rabi_2006}, see also Appendix~B.1. in \cite{kaufmann_scalable_2017b}). 
Here, $g_{\mathrm{c},nm,\kappa\lambda} = \sqrt{g_{1,nm,\kappa\lambda}^2 + g_{2, nm,\kappa\lambda}^2}$, and the coefficients
\begin{align}
g_{1, nm, \kappa\lambda} &= \frac{\sqrt{2}}{2}\, \Omega_{nm, \kappa\lambda}\\ 
g_{2, nm, \kappa\lambda} & = \frac{\sqrt{2}}{2}\, \Omega_{n+\kappa\,m+\lambda, \kappa\lambda}
\end{align}
are the couplings between the states $\ket{\downarrow\downarrow} \leftrightarrow \frac{1}{\sqrt{2}}(\ket{\uparrow\downarrow} + \ket{\downarrow\uparrow})$ and the states $\frac{1}{\sqrt{2}}(\ket{\uparrow\downarrow} + \ket{\downarrow\uparrow}) \leftrightarrow \ket{\uparrow\uparrow} $, respectively. 
To include the phenomenological decay, we modify the equations above in an analogous way to the case of a single ion to 
\begin{align}
\begin{split}
|c^{(0)}_{nm,\kappa\lambda}|^2 = & \frac{1}{g_{\mathrm{c},nm,\kappa\lambda}^4}\big[g_{1,nm,\kappa\lambda}^{4} / 2 \\
&\;\;\; + g_{1,nm,\kappa\lambda}^{4} / 2 \cdot  \cos \left(2 g_{\mathrm{c},nm,\kappa\lambda} t\right) \exp(-\gamma t) \\ 
	&\;\;\; +2 g_{1,nm,\kappa\lambda}^{2} g_{2,nm,\kappa\lambda}^{2}\cos(g_{\mathrm{c},nm,\kappa\lambda} t)\exp(-\gamma t)  \\ 
	& \;\;\; +g_{2,nm,\kappa\lambda}^{4}\big] \\
|c^{(1)}_{nm,\kappa\lambda}|^2 = & \frac{g_{1,nm,\kappa\lambda}^2}{g_{\mathrm{c},nm,\kappa\lambda}^2} \big[1/2 \\
& \;\;\;-1/2 \cdot \cos \left(2g_{\mathrm{c},nm,\kappa\lambda} t\right) \exp(-\gamma t)\big] 
\,,
\end{split}
\end{align}
with the decay terms added to the oscillating terms of the expected population.

Transport could potentially produce a coherent state of motion when coherent excitation is not properly removed at the end of the transport \cite{bowler_coherent_2012}. Therefore, we also analyze the data assuming a coherent distribution of motion and derive an average occupation number $\bar{n}$ for each measurement. In general, this yields a smaller $\bar{n}$ compared to the case where we assume a thermal distribution. For small $\bar{n}$ ($\bar{n}\approx 0.1$), the $\bar{n}$ derived from the two analyses have overlapping 1-$\sigma$ confidence intervals, since the two distributions are very similar in this regime. For large $\bar{n}$ ($\bar{n} > 1$), the $\bar{n}$ derived from the two analyses have non-overlapping 3-$\sigma$ confidence intervals. However, the reduced-$\chi^2$ from analysis using a coherent distribution shows a larger deviation from $1$ compared to the number derived using a thermal distribution, indicating that a thermal distribution is  more appropriate here. As a conservative description for all regimes, we present the results derived using a thermal distribution in the main text.


\subsection{Mode rotation in junction}
To shuttle a single \be ion from \szone to \vzone, the orientation of the axis of the weakest confinement is changed from being parallel to the $z$-axis to being parallel to the $y$-axis. 
This can lead to degeneracy between the near-groundstate-cooled mode and the  radial modes cooled to near the Doppler limit during the transport process \cite{blakestad_near-ground-state_2011}. 
To avoid mode energy exchange near the junction, the potential well at \czone is adiabatically rotated by $90^\circ$ (labeled as \cyzone) from being aligned to the $z$-axis to being aligned to the $x$-axis before being moved further up to \vzone \cite{splatt_deterministic_2009,kaufmann_fast_2017}.

\end{document}